\documentclass[floatfix,showpacs,pre,superscriptaddress]{revtex4}
\usepackage{graphicx}
\usepackage{amsmath}
\usepackage{amsfonts}
\usepackage{amssymb}

\newcommand{\be}{\begin{eqnarray}}
\newcommand{\ee}{\end{eqnarray}}
\newcommand{\vS}{\mathbf{S}}
\newcommand{\p}{\partial}

\def\ket#1{|#1\rangle}

\def\ep#1{\langle #1 \rangle}
\def\z#1{z_{x_{#1}}}

\begin{document}

\title{Nonlocal behaviors of spin correlations in the Haldane-Shastry model}

\author{Maowen Tang}
\affiliation{College of Physics, Sichuan University, 610064,
Chengdu, P. R. China}

\author{Yan He}
\email{heyan_ctp@scu.edu.cn}
\affiliation{College of Physics, Sichuan University, 610064,
Chengdu, P. R. China}

\date{\today}

\begin{abstract}
The nonlocal factors of spin correlations are introduced for lattice spin models. Based on this concept, we investigate the nonlocal behavior of the Haldane-Shastry model with or without ring frustration. The ground state and spin correlations of the Haldane-Shastry model are calculated for both even and odd number of spins, then the nonlocal factors can be deduced analytically. It is found that the nonlocal factor due the ring frustration is the same as the Heisenberg model.
\end{abstract}

\maketitle

\section{Introduction}

It is known for a long time that the one dimensional spin models with odd number of total spin may display some special features different from the one with even number of spins \cite{Bariev,Cabrera,CabreraPRB,Karbach,Cador,Barwinkel,Baker}. Recently, there are rising interests in the effects of ring frustration in spin systems with anti-ferromagnetic couplings and odd number of spins \cite{Campostrini,CampostriniJSM,DongJSM}. The ring frustration is a type of geometrical frustration caused by the impossibility of accommodating the staggered spin configuration on a closed spin chain with odd number of sites. In the transverse Ising model, it is found that there appears some kink like excitations due to this frustration, which makes the dispersion gapless rather than gapful \cite{DongJSM} and also modifies the thermodynamic quantities at finite temperature \cite{He}. Previous work \cite{DongPRE} also showed that the spin correlations display some nonlocal behaviors due to the ring frustration. Besides the transverse Ising model, the nonlocal factors of correlations can also be obtained numerically for both XY model and isotropic Heisenberg model \cite{LiPRE}. In this paper, we will further investigate the nonlocal behavior of the Haldane-Shastry (HS) model\cite{Haldane,Shastry}, a spin model with inverse square exchange couplings. Before the discussion of HS model, we will first briefly recall the definition of nonlocal factors.

The spin correlation function of the ground state $|\Psi\rangle$ is given by
\be
C^a_{r,N}=\ep{\Psi|S_{j}^{a}S_{j+r}^{a}|\Psi},
\label{Correlator}
\ee
where $S_j^a=\frac12\sigma_{j}^a$ and $\sigma^a$ with $a=x,y,z$are Pauli matrices. Here the $r$ is the distance between the two spins and $N$ is the total number of spins. In this paper, we only consider the $x$ component of the correlation function $C^x_{r,N}$. To ease the notation, we will drop the superscript and simply denote it as $C_{r,N}$. Obviously, the correlations should satisfy the following cyclic relation
\be
C_{r,N}=C_{N-r,N}.
\label{cyclic}
\ee

In field theory, one usually consider the local behavior corresponding to the case $r\ll N$. In this case, one can take the continuum limit of the correlations obtained from a lattice model, such as replacing the sum of momentum $q$ with an integral (in $D$ dimensions), $(1/N)\sum_{q}=\int\mathrm{d}^{D}q/(2\pi)^D$. Thus the system size $N$ drops off the final results after taking the continuum limit. We denote the resulting correlation function as $C_{\infty}(r)$. For example, the correlations of critical spin chains have been found to exhibit algebraically decaying correlation functions \cite{CFT-book} like
\begin{equation}
C_{\infty}(r)\sim\frac{b}{r^\eta},
\label{algebraic}
\end{equation}
where $b$ and $\eta$ are some constants. In one dimension, there may also be some logarithm corrections to the above correlations.

One the other hand, we can also consider the nonlocal behavior of correlations by taking both $r\to\infty$ and $N\to\infty$ but keeping the ratio $\alpha=r/N$ fixed. In this case, we can define the following correlation functions
\be
C^{(O)}(r,\alpha)=\lim_{L\rightarrow\infty}C_{r,2L+1},\qquad
C^{(E)}(r,\alpha)=\lim_{L\rightarrow\infty}C_{r,2L},
\ee
for odd number of spins $N=2L+1$ and even number of spins $N=2L$ respectively. Note that $0\leq\alpha<1/2$ due to the ring geometry and the cyclic relation of Eq. (\ref{cyclic}). Because of the ring frustration, the nonlocal behaviors of $C^{(O)}(r, \alpha)$ and $C^{(E)}(r, \alpha)$ are quite different and we have to treat them separately. This feature usually cannot be easily captured by the continuum field theory.

The ratios between the nonlocal correlation function $C^{(O)}(r, \alpha)$, $C^{(E)}(r, \alpha)$ and the local correlation function $C_{\infty}(r)$ define the nonlocal factors. The first two of them are for the measure of nonlocality of the system with even or odd number of spins respectively,
\be
R^{(O)}(\alpha)=\frac{C^{(O)}(r, \alpha)}{C_{\infty}(r)},\qquad
R^{(E)}(\alpha)=\frac{C^{(E)}(r, \alpha)}{C_{\infty}(r)}.
\ee
In these definitions, we assume that the $r$ dependence of the right hand sides of above equations cancels out.
The last one measures the effect of ring frustration,
\be
R(\alpha)=\frac{R^{(O)}(\alpha)}{R^{(E)}(\alpha)}.
\label{R}
\ee

Actually, the nonlocal factor has already been widely studied under the context of finite-size scaling (FSS) hypothesis \cite{Kaplan}. But the effects of the ring frustration has not been quite emphasized before. Now we give a few examples of nonlocal factors of different types of spin models. The nonlocal factors of spin correlations of transverse Ising model at critical point can be derived analytically \cite{LiPRE}, and the results are
\be
R^{(E)}(\alpha)=\exp\Big[\frac{\alpha}2-\frac{\pi^2\alpha^2}{24}+\cdots\Big],\qquad
R(\alpha)=\cos\frac{\pi\alpha}2-\sin\frac{\pi\alpha}2
\ee
For a similar model like XY model, both nonlocal factors can only be obtained by numerical regression, and the results are ($R^{(E)}$ of XY model was already obtained in \cite{Kaplan})
\be
R^{(E)}(\alpha)=1+0.28822\sinh^2(1.673\alpha),\qquad
R(\alpha)=\cos(\pi\alpha)
\ee
For a more complicated model such as Heisenberg model, the correlation function can only be numerically computed up to 30 spins either by exact diagonalization or by exact solution. Although the data size is quite small, one can still numerically deduce the nonlocal factors as ($R^{(E)}$ of Heisenberg model was already obtained in \cite{Hallberg})
\be
R^{(E)}(\alpha)=\Big[1+0.28822\sinh^2(1.673\alpha)\Big]^{1.805},\quad
R(\alpha)=\cos(\pi\alpha)
\ee
As $\alpha\to 0$ one should get back to the local behaviors. Therefore we have $R^{(E)}(0)=R(0)=1$ in all the above models as they should be. The effect of ring frustration is manifested in that $R(1/2)=0$. This is true for all the three different spin models. Since the HS model has a long range interacting between spins, one would expect its ground state will be more disordered than other spin models. But we will see that actually $R(\alpha)$ of HS model is the same as that of Heisenberg model. The advantage of HS model is that its ground state has much simpler functional form. Therefore, its the nonlocal factors of both even and odd case can be almost analytically calculated, as we will explain in details in the rest of this paper.

\section{The ground state of Haldane-Shastry model}

The HS model is given by
\be
H=\sum_{i<j}\frac{J}{[D(i-j)]^2}\vS_i\cdot\vS_j,\quad D(i-j)=\frac{\pi}{N}\sin^2\Big[\frac{(i-j)}{N}\pi\Big]
\ee
where $S_j^a=\frac12\sigma_{j}^a$ and $\sigma^a$ with $a=x,y,z$ are Pauli matrices. The system has $N$ lattice sites with unit lattice constant. We assume periodic boundary condition, thus the $N$ spins form a ring and $D(i-j)$ is the chord distance between spin $\vS_i$ and $\vS_j$. The HS model has a coupling proportional to the inverse square of the distance.

A general state of a $N$ spin system can always be written as
\be
\ket{\Psi}=\sum_{x_1<\cdots<x_M}\Psi(x_1,\cdots,x_M)S^-_{x_1}\cdots S^-_{x_M}\ket{0}
\ee
where $\ket{0}=\ket{\uparrow\cdots\uparrow}$ is the reference state with all spins being spin up. We introduce the spin lowering operator $S_j^-=S^x_j-iS^y_j$ which will create a down spin at site $j$. The coordinates $x_1,\cdots,x_M \in (1,\cdots,N)$ give the locations of the $M$ down spins. The summation of $x_j$ means summing over all possible locations $\sum_{x_j}=\sum_{x_j=1}^N$. Note that $S^z=\sum_jS^z_j$ commutes with the Hamiltonian, thus the number of down spins is conserved. Therefore each Hilbert subspace with different number of down spins $M$ can be considered separately. For even $N$, the ground state occurs in the subspace with $M=N/2$. For odd $N$, the ground state occurs in the subspace with $M=(N\pm1)/2$.

The ground state wave function of HS model is known to have the Gutzwiller-Jastrow type of function form \cite{HS-book}. It can be thought as the ground state of non-interacting spin one half fermions system with single occupation at each site. For both even and odd $N$, the ground state wave function can be expressed as
\be
\Psi(x_1,\cdots,x_M)= \prod_{i=1}^{M}\z{i}\prod_{1\leq i<j\leq M}(\z{i}-\z{j})^2
\ee
with $\z{j}=\exp\Big(i\frac{2\pi x_j}{N}\Big)$.

Now we briefly describe how to derive the above wave function for odd $N$. Due to the periodic boundary condition, the lattice momentum is $k_i=\frac{2\pi}{N}n_i$ with $n_i=-\frac{N-1}{2},\cdots,\frac{N-1}{2}$. For concreteness, we assume that $M=(N-1)/2$ and $M$ is odd. Then the spin up fermions will occupy the states with the $M$ smallest lattice momenta $k_i$, which correspond to $n_i=-\frac{M-1}{2},\cdots,\frac{M-1}{2}$. The wave function of spin up fermions is given by the Slater determinant
\be
\Psi_{\uparrow}(x_1,\cdots,x_M)=\mbox{det}\Big[\exp(i k_ix_j)\Big]_{M\times M}
=\prod_{i=1}^{M}\z{i}^{-(M-1)/2}\prod_{1\leq i<j\leq M}(\z{i}-\z{j})
\ee
For the spin down fermions, we will use the hole representation for convenience. To satisfy the single occupation condition, one only need to require that the holes of the spin down fermions locate at the same positions as the particles of spin up fermions. Then the holes of spin down fermions will occupy the states with the $M$ largest lattice momenta $k_i$, which correspond to $n_i=\frac{M+1}{2},\cdots,\frac{3M-1}{2}$. (Note that $n_i=\frac{3M-1}{2}$ is equivalent to $n_i=-\frac{M+3}{2}$, since the lattice momentum space is also periodic.) The wave function of the holes of spin down fermions is given by
\be
\Psi_{\downarrow}(x_1,\cdots,x_M)=\mbox{det}\Big[\exp(i k_ix_j)\Big]_{M\times M}
=\prod_{i=1}^{M}\z{i}^{(M+1)/2}\prod_{1\leq i<j\leq M}(\z{i}-\z{j})
\ee
Then the ground state wave function of HS model is given by
\be
\Psi(x_1,\cdots,x_M)=\Psi_{\uparrow}\Psi_{\downarrow}
=\prod_{i=1}^{M}\z{i}\prod_{1\leq i<j\leq M}(\z{i}-\z{j})^2
\label{g1}
\ee
which is the same as the ground state wave function when $N$ is even.

For the wave function of holes, there is another choice of the $M$ occupied states, which correspond to $n_i=\frac{M+3}{2},\cdots,\frac{3M+1}{2}$. (Note that $n_i=\frac{3M+1}{2}$ is equivalent to $n_i=-\frac{M+1}{2}$.)
In this case, the hole wave function is given by
\be
\Psi_{\downarrow}(x_1,\cdots,x_M)=\mbox{det}\Big[\exp(i k_ix_j)\Big]_{M\times M}
=\prod_{i=1}^{M}\z{i}^{(M+3)/2}\prod_{1\leq i<j\leq M}(\z{i}-\z{j})
\ee
And the ground state wave function becomes
\be
\Psi(x_1,\cdots,x_M)=\Psi_{\uparrow}\Psi_{\downarrow}
=\prod_{i=1}^{M}\z{i}^2\prod_{1\leq i<j\leq M}(\z{i}-\z{j})^2
\label{g2}
\ee
If $M$ is even, we can follow the almost same steps to derive the ground state Eq.(\ref{g1}) and (\ref{g2}). If $M=(N+1)/2$, the ground state can be obtained from the case of $M=(N-1)/2$ by switching the up and down spin. Altogether, for odd $N$, there are 4 degenerate ground state corresponding to $M=(N\pm1)/2$ and Eq.(\ref{g1}),(\ref{g2}). We would like to mention that the Heisenberg model with odd number of spins also has 4 degenerate ground states.

\section{The spin correlation of Haldane-Shastry model}

In this section, we will derive the spin correlation function of HS model for both even and odd number of spins with periodic boundary condition. Then we will deduce its nonlocal factors. The spin correlation is
\be
C_{lm}=\frac{\ep{\Psi|S_l^x S_m^x|\Psi}}{\ep{\Psi|\Psi}}
\ee
where $\ket{\Psi}$ is unnormalized the ground state of HS model. It can also be expressed by
\be
C_{lm}=\frac12(C_{lm}^{+-}+C_{lm}^{-+}),\qquad C_{lm}^{-+}=\frac{\ep{\Psi|S_l^- S_m^+|\Psi}}{\ep{\Psi|\Psi}}
\ee
where $S^{\pm}=S_x\pm iS_y$ are ladder operators.

Since the ground state Eq.(\ref{g1}) and (\ref{g2}) are not normalized, we have to compute the norm of the ground state first. The basic idea of this calculation is to rewrite the norm in terms of a special determinant called confluent alternant. Then expand the determinant by Laplace theorem and make use the fact that the summation of nonzero power of unit roots is zero, one will arrive at the final result. The details is shown in appendix \ref{app-norm}. Similar technique will also be used to compute the numerator of $C^{+-}_{lm}$.

Applying the raising operator to the ground state, we find that
\be
S_{m}^{+}\ket{\Psi}=\sum_{x_{1}<\ldots <x_{M-1}}\Psi(m,x_{1},\ldots ,x_{M-1})S_{x_{1}}^{-}\cdots S_{x_{M-1}}^{-}\ket{0}
\ee
The numerator of $C^{-+}_{lm}$ is the inner product of the above states, which can be written as
\be
\nonumber
&&\ep{\Psi | S_{l}^{-}S_{m}^{+} |\Psi}=\sum_{x_{1}<\ldots <x_{M-1}}
\Psi^{\ast}(l,x_{1},\ldots ,x_{M-1})\Psi(m,x_{1},\ldots,x_{M-1})\nonumber\\
&&=\frac{1}{(M-1)!}\sum_{x_{1},\ldots ,x_{M-1}}z_{l}^{\ast}z_{m}
\prod_{i=1}^{M-1}\z{i}^{\ast}\z{i}(z_{l}^{\ast}-\z{i}^{\ast})^{2}(z_{m}-\z{i})^{2}
\prod_{1\leq i<j\leq M-1}(\z{i}^{\ast}-\z{j}^{\ast})^{2}(\z{i}-\z{j})^{2}\nonumber\\
&&=\frac{1}{(M-1)!}f_{lm}\, S
\ee
In the last step, $z_j^*=1/z_j$ is used. For convenience, we have introduced
\be
&&f_{lm}=\left\{
           \begin{array}{ll}
             z_l^2z_m, & \hbox{for} N=2M+1\\
             z_l z_m, & \hbox{for} N=2M
           \end{array}
         \right.\\
&&S=\sum_{x_{1},\ldots ,x_{M-1}}\prod_{i=1}^{M-1}\z{i}^{-2(M-1)}
\Big[\prod_{i=1}^{M-1}(z_{l}-\z{i})^{2}(z_{m}-\z{i})^{2}
\prod_{1\leq i<j\leq M-1}(\z{i}-\z{j})^{4}\Big]
\ee
The factors inside the bracket can be rewritten in terms of a confluent alternant with the help of Eq.(\ref{alter1}), thus we find
\be
S=-\frac{1}{z_{l}-z_{m}}\sum_{x_{1},\ldots ,x_{M-1}}\z{1}^{-2M+2}\cdots\z{M-1}^{-2M+2}\mbox{det}
\begin{pmatrix}
1 & z_{l} & z_{l}^{2} & z_{l}^{3} & \cdots & z_{l}^{2M-1} \\
1 & z_{m} & z_{m}^{2} & z_{m}^{3} & \cdots & z_{m}^{2M-1} \\
1 & \z{1} & \z{1}^{2} & \z{1}^{3} & \cdots & \z{1}^{2M-1} \\
0 & 1 & 2\z{1} & 3\z{1}^{2} & \cdots & (2M-1)\z{1}^{2M-2} \\
\vdots & \vdots & \vdots & \vdots & \ddots & \vdots \\
1 & \z{M-1} & \z{M-1}^{2} & \z{M-1}^{3} & \cdots & (2M-1)\z{M-1}^{2M-2}
\end{pmatrix}
\ee
Similar to the calculation of the norm, the confluent alternant can be expanded by any two rows through Laplace theorem. Most terms will vanish due to the summation $\sum_{x_j}\z{j}^n=\delta_{n0}$. The only nonzero contributions come from the terms that all $\z{j}$ has zero power. Therefore, we find that
\be
\nonumber
S&=& -\frac{1}{z_{l}-z_{m}}\Big [N^{M-1}(M-1)!\frac{(2M-1)(2M-3\cdots 1)}{2M-1}\mbox{det}
\begin{pmatrix}
1 & z_{l}^{2M-1} \\
1 & z_{m}^{2M-1}
\end{pmatrix} \\ \nonumber
& & +N^{M-1}(M-1)!\frac{(2M-1)(2M-3)\cdots 1}{2M-3}\mbox{det}
\begin{pmatrix}
z_{l} & z_{l}^{2M-2} \\
z_{m} & z_{m}^{2M-2}
\end{pmatrix}
+\cdots\Big ] \\ \nonumber
&=& -\frac{1}{z_{l}-z_{m}}\frac{N^{M-1}(M-1)!(2M!)}{2^{M}M!}\Big [\frac{1}{2M-1}(z_{m}^{2M-1}-z_{l}^{2M-1})+\frac{1}{2M-3}(z_{l}z_{m}^{2M-2}-z_{l}^{2M-2}z_{m})+\cdots\Big ] \\
&=& -\frac{1}{z_{l}-z_{m}}\frac{N^{M-1}(2M!)}{2^{M}M}\sum_{p=1}^{M}\Big [\frac{1}{2p-1}(z_{l}^{M-p}z_{m}^{M+p-1}-z_{l}^{M+p-1}z_{m}^{M-p})\Big ]
\ee
Now collect all the above results and also make use of the norm of Eq.(\ref{norm}), we find that
\be
C_{lm}^{-+} &=& \frac{f_{lm}\, S}{(M-1)!}\frac{1}{\ep{\Psi|\Psi}}=-\frac{1}{N}\frac{f_{lm}}{z_{l}-z_{m}}\sum_{p=1}^{M}\Big [\frac{1}{2p-1}(z_{l}^{M-p}z_{m}^{M+p-1}-z_{l}^{M+p-1}z_{m}^{M-p})\Big]\nonumber\\
&=&-\frac{1}{N}\frac{1}{z_{l}-z_{m}}\Big(\frac{z_m}{z_l}\Big)^M\sum_{p=1}^{M}
\Big[\frac{1}{2p-1}(z_{l}^{1-p}z_{m}^{p}-z_{l}^{p}z_{m}^{1-p})\Big]
\label{Cpm}
\ee
Here we have used the fact that $z_l^N=1$. The above result is correct for both even and odd number of $N$.

We can also compute the spin correlation with the following degenerate ground state wave function
\be
\ket{\Psi} = \prod_{i=1}^{M}\z{i}^{2}\prod_{1\leq i<j\leq M}(\z{i}-\z{j})^2
\ee
Following very similar steps, we find that the spin correlations is given by
\be
\nonumber
C_{lm}^{-+} &=& -\frac{1}{N}\frac{z_{l}z_{m}^{2}}{z_{l}-z_{m}}\sum_{p=1}^{M}\Big [\frac{1}{2p-1}(z_{l}^{M-p}z_{m}^{M+p-1}-z_{l}^{M+p-1}z_{m}^{M-p})\Big ] \\
&=&-\frac{1}{N}\frac{1}{z_{l}-z_{m}}\Big(\frac{z_l}{z_m}\Big)^M\sum_{p=1}^{M}
\Big[\frac{1}{2p-1}(z_{l}^{1-p}z_{m}^{p}-z_{l}^{p}z_{m}^{1-p})\Big]
\ee
which will lead to the same $C_{lm}$ as Eq.(\ref{Cpm}). Therefore, the 4 degenerate ground states give rise to the same spin correlations.

\begin{figure}
\centerline{\includegraphics[width=0.9\textwidth]{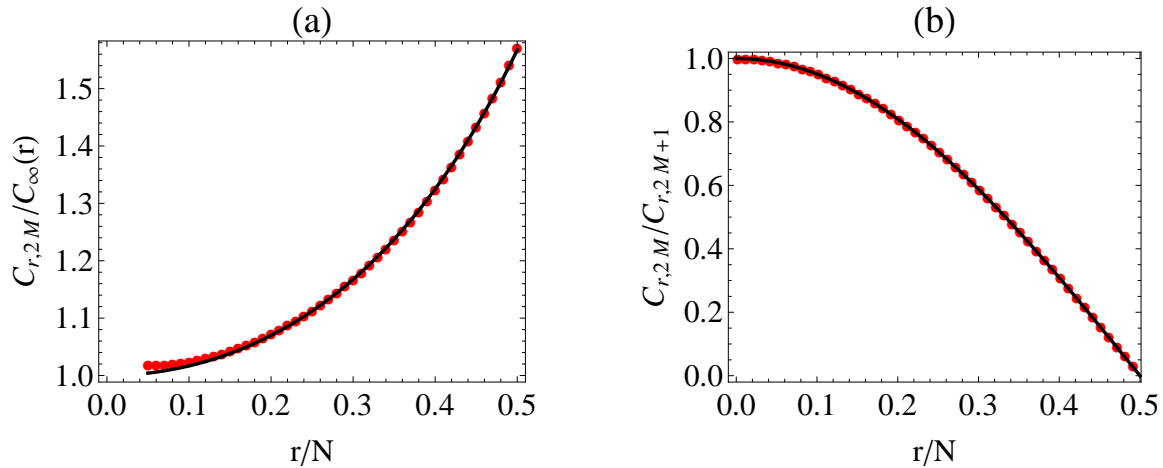}}
\caption{In panel (a), the red dots show $C_{r,2M}/C_{\infty}(r)$ as a function of $r/N$ for M=1000. The black curve is the fitting curve. In panel (b), the red dots show $C_{r,2M+1}/C_{r,2M}$ as a function of $r/N$ for M=500. The black curve is $R(\alpha)$ of Eq.(\ref{Ra}) }
\label{Corr}
\end{figure}

\section{The Nonlocal factors of Haldane-Shastry Model}

Now we turn to the discussion of nonlocal factors. We first consider the case with even number of spins $N=2M$. In this case $(z_m/z_l)^M=\pm1$. Since we only care about the magnitude of the correlations, this factor does not matter. Make use of the fact that $z_l=\exp(i\theta_l)$ with $\theta_l=2\pi x_l/N$, we find the following identity
\be
\frac{z_{l}^{1-p}z_{m}^{p}-z_{l}^{p}z_{m}^{1-p}}{z_l-z_m}
=-\frac{\sin{(p-\frac{1}{2})(\theta_{l}-\theta_{m})}}{\sin{\frac{\theta_{l}-\theta_{m}}{2}}}
\ee
Then the spin correlation is
\be
C_{r,2M}=\mbox{Re}\,C^{-+}_{lm}=\frac{1}{2M\sin(\theta/2)}\sum_{p=1}^M\frac{1}{2p-1}\sin\frac{(2p-1)\theta}2
\label{CE}
\ee
with $\theta=2\pi r/(2M)$ and $r=x_l-x_m$. This spin correlation of HS model has a simpler function form comparing to the transverse field Ising model whose correlations can be expressed as a Toeplitz determinant \cite{McCoy}.

Making use the following identity
\be
\sum_{p=1}^M\cos\frac{(2p-1)\theta}2=\frac{\sin M\theta}{2\sin(\theta/2)}
\ee
we can transfer the spin correlation of Eq.(\ref{CE}) into an integral form as
\be
C_{r,2M}=\frac{1}{4M\sin(\theta/2)}\int_0^{\theta}dt\frac{\sin M t}{2\sin(t/2)}
=\frac{1}{4M\sin(\theta/2)}\int_0^{\pi r}dt\frac{\sin t}{2M\sin\frac{t}{2M}}
\label{CM}
\ee
Suppose that $r\ll 2M$, we find the the continuum limit of spin correlations as
\be
C_{\infty}(r)=\lim_{M\to\infty}C_{r,2M}=\frac{1}{2\pi r}\int_0^{\pi r}dt\frac{\sin t}{t}
\ee
In Figure \ref{Corr} panel (a), we plot the ratio $C_{r,2M}/C_{\infty}(r)$  (red dots) as a function of $\alpha=r/N$ in the range $0<\alpha<1/2$. Due to the cyclic relation Eq.(\ref{cyclic}), $C_{r,2M}$ is symmetric about $\alpha=1/2$, thus there is no need to go beyond this point. The data of red dots can be fitted into the black curve given by the following function form
\be
R^{(E)}(\alpha)=1+0.428\sinh^2(1.969\alpha)
\label{RE1}
\ee
This is the nonlocal factor of HS model for even number of spins. We can also analytically derive an approximate expression for this nonlocal factor. First the integral in Eq.(\ref{CM}) can be expanded as
\be
\int_0^{\pi r}dt\frac{\sin t}{2M\sin\frac{t}{2M}}
=\int_0^{\pi r}dt\Big[\frac{\sin t}{t}+\frac{t\sin t}{24M^2}\cdots\Big]
=\int_0^{\pi r}dt\frac{\sin t}{t}-\frac{-\pi\alpha\cos(\pi r)}{M}+\cdots
\ee
As $M\to\infty$, we expect that the first order and higher order terms all approach zero. Therefore, we find
\be
C_{r,2M}\approx\frac{\alpha}{2r\sin(\pi\alpha)}\int_0^{\pi r}dt\frac{\sin t}{t}
\ee
From this form of correlation, we can extract the nonlocal factor as
\be
R^{(E)}(\alpha)=\frac{\pi\alpha}{\sin(\pi\alpha)}
\ee
which gives an almost identical curve as the numerical fitting curve of Eq.(\ref{RE1}).

Now we turn to the case with odd number of spins $N=2M+1$. In this case, the spin correlation is given by
\be
C_{r,2M+1}=\mbox{Re}\,C^{-+}_{lm}=\frac{\cos (M\theta')}{(2M+1)\sin(\theta'/2)}
\sum_{p=1}^M\frac{1}{2p-1}\sin\frac{(2p-1)\theta'}2
\ee
with $\theta'=2\pi r/(2M+1)$ and $r=x_l-x_m$. It is easy to verify that the extra factor can be written as
\be
\cos(M\theta')=\cos\Big(M\frac{2\pi r}{2M+1}\Big)=(-1)^r\cos\Big(\pi\frac{r}{2M+1}\Big)
\ee
Set aside the unimportant minus sign, and note that $\theta'\approx\theta$ for very large $M$, we find that
\be
C_{r,2M+1}\approx \cos\Big(\pi\frac{r}{2M+1}\Big)C_{r,2M}
\ee
Therefore, we identity the nonlocal factor due to ring frustration as
\be
R(\alpha)=\cos(\pi\alpha). \label{Ra}
\ee
which is exact the same as the Heisenberg model.

\section{Conclusion}

In this paper, we have shown that the spin correlation of HS model can be factorized into two parts. One part is the local form $C_{\infty}(r)$ which is obtained by taking continuum limit of the lattice model. The other one is the nonlocal factor $R^{(E)}(\alpha)$ and $R(\alpha)$ which account the nonlocal behaviors of correlations for even and odd number of total spins. This is consistent with our previous proposal \cite{LiPRE} of factorizable correlations of spin models. Actually, the HS model provide us a much more calculable example. Although the HS model has a long range interaction, its ground state is much simpler than the Heisenberg model. For the later case, the Bethe ansatz ground state involves a summation of permutations of $M$ down spins, which makes the analytical calculation of spin correlations too complicated. One the other hand, the simple correlations of HS model allow us to deduce its nonlocal factors semi-analytically. The ring frustration manifests itself as the same nonlocal factor $R(\alpha)$ for both HS model and Heisenberg model. From this result, we suspect that there might be some universal properties for spin model with odd number of spins. To reveal them requires further investigations.

\acknowledgments
We are indebted to Hong-Hao Tu for providing us his note on HS model. Without his note, this work is impossible. We also thank the useful discussions with Peng Li. This work is supported by NSFC under Grant No. 11874272.

\appendix

\section{The calculation of the norm of ground state wave function}
\label{app-norm}

The ground state of HS model can be written as
\be
\ket{\Psi}=\sum_{x_1<\cdots<x_M}\Psi(x_1,\cdots,x_M)S^-_{x_1}\cdots S^-_{x_M}\ket{0}
\ee
where $\ket{0}=\ket{\uparrow\cdots\uparrow}$ is the reference state with all up spins. Here $x_1,\cdots,x_M \in (1,\cdots,N)$ and $N$ is the system size. The summation means sum over all possible values
$\sum_{x_j}=\sum_{x_j=1}^N$.
The wave function can be expressed as
\be
\Psi(x_1,\cdots,x_M)= \prod_{i=1}^{M}\z{i}\prod_{1\leq i<j\leq M}(\z{i}-\z{j})^2
\ee
with $\z{j}=\exp\Big(i\frac{2\pi x_j}{N}\Big)$.

The norm square of the ground state can be written as
\be
\ep{\Psi|\Psi}=\sum_{x_{1}<\ldots <x_{M}}|\Psi(x_{1},\ldots ,x_{M})|^2
=\frac{1}{M!}\sum_{x_{1},\ldots ,x_{M}}|\Psi(x_{1},\ldots ,x_{M})|^2
\ee
Note that $|\z{j}|^2=1$, then the norm can be rewritten as
\be
\nonumber
&&|\Psi(x_{1},\ldots ,x_{M})|^2=\prod_{1\leq i<j\leq M}|\z{i}-\z{j}|^4
=\prod_{1\leq i<j\leq M}(\z{i}-\z{j})^2(\z{i}^{\ast}-\z{j}^{\ast})^2 \\
&&=\prod_{1\leq i<j\leq M}\frac{(\z{i}-\z{j})^4}{\z{i}^2\z{j}^2}
= \prod_{l}\z{l}^{-2(M-1)}\prod_{1\leq i<j\leq M}(\z{i}-\z{j})^4
\label{z4}
\ee
The second factor of Eq.(\ref{z4}) can be expressed as an determinant called ``confluent alternant'' through the identity of Eq.(\ref{alter}), thus we find
\be
\nonumber
\ep{\Psi|\Psi} &=& \frac{1}{M!}\sum_{x_{1},\ldots ,x_{M}}\z{1}^{-2M+2}\cdots\z{M}^{-2M+2}\mbox{det}
\begin{pmatrix}
1 & \z{1} & \z{1}^2 & \z{1}^3 & \cdots & \z{1}^{2M-1} \\
0 & 1 & 2\z{1} & 3\z{1}^2 & \cdots & (2M-1)\z{1}^{2M-2} \\
1 & \z{2} & \z{2}^2 & \z{2}^3 & \cdots & \z{2}^{2M-1} \\
0 & 1 & 2\z{2} & 3\z{2}^2 & \cdots & (2M-1)\z{2}^{2M-2} \\
\vdots & \vdots & \vdots & \vdots & \ddots & \vdots \\
1 & \z{M} & \z{M}^2 & \z{M}^3 & \cdots & \z{M}^{2M-1}\\
0 & 1 & 2\z{M} & 3\z{M}^2 & \cdots & (2M-1)\z{M}^{2M-2}
\end{pmatrix} \\ \nonumber
\ee
The confluent alternant can be expanded by any two rows through Laplace theorem. Note that $\sum_{x_j}\z{j}^n=\delta_{n0}$. Because of this summation, most of the terms in the Laplace expansion is zero. The only nonzero contribution comes from the terms that all $\z{j}$ has zero power. Therefore, we find the following result.
\be
\ep{\Psi|\Psi}&=& \frac{1}{M!}\sum_{x_{1},\ldots ,x_{M}}\z{1}^{-2M+2}\cdots\z{M}^{-2M+2}\Big [\mbox{det}
\begin{pmatrix} 1 & \z{1}^{2M-1}\\
0 & (2M-1)\z{1}^{2M-2}
\end{pmatrix}
\mbox{det}(\cdots)\nonumber\\
& &+\mbox{det}
\begin{pmatrix} \z{1} & \z{1}^{2M-2}\\
1 & (2M-2)\z{1}^{2M-3}
\end{pmatrix}
\mbox{det}(\cdots)+\cdots\Big] \nonumber\\
&=& \frac{1}{M!}N^{M}M![(2M-1)(2M-3)\cdots 1]= N^M\frac{(2M)!}{2^M M!}
\label{norm}
\ee
This result is correct for both $N=2M$ and $N=2M+1$.

Clearly, the calculation will be the same if we use another form of wave function
\be
\ket{\Psi} = \prod_{i=1}^{M}\z{i}^{2}\prod_{1\leq i<j\leq M}(\z{i}-\z{j})^2
\ee
The norm of the above wave function is the same as Eq.(\ref{norm}).

\section{Confluent alternant}

In this part, we derive two useful identities of the special determinant called confluent alternant.
\be
&&\mbox{det}
\begin{pmatrix}
1 & z_{1} & z_{1}^{2} & z_{1}^{3} & \cdots & z_{1}^{2M-1} \\
0 & 1 & 2z_{1} & 3z_{1}^{2} & \cdots & (2M-1)z_{1}^{2M-2} \\
1 & z_{2} & z_{2}^{2} & z_{2}^{3} & \cdots & z_{2}^{2M-1} \\
0 & 1 & 2z_{2} & 3z_{2}^{2} & \cdots & (2M-1)z_{2}^{2M-2} \\
\vdots & \vdots & \vdots & \vdots & \ddots & \vdots \\
1 & z_{M} & z_{M}^{2} & z_{M}^{3} & \cdots & z_{M}^{2M-1} \\
0 & 1 & 2z_{M} & 3z_{M}^{2} & \cdots & (2M-1)z_{M}^{2M-2}
\end{pmatrix}=\prod_{1\leq i<j\leq M}(z_{i}-z_{j})^{4}
\label{alter}
\ee
\be
&&\mbox{det}
\begin{pmatrix}
1 & y_1 & y_1^{2} & y_1^{3} & \cdots & y_1^{2M-1} \\
1 & y_2 & y_2^{2} & y_2^{3} & \cdots & y_2^{2M-1} \\
1 & z_1 & z_1^{2} & z_1^{3} & \cdots & z_1^{2M-1} \\
0 & 1 & 2z_1 & 3z_1^{2} & \cdots & (2M-1)z_1^{2M-2} \\
\vdots & \vdots & \vdots & \vdots & \ddots & \vdots \\
1 & z_{M-1} & z_{M-1}^{2} & z_{M-1}^{3} & \cdots & z_{M-1}^{2M-1}\\
0 & 1 & 2z_{M-1} & 3z_{M-1}^{2} & \cdots & (2M-1)z_{M-1}^{2M-1}
\end{pmatrix}\nonumber\\
&&= -(y_1-y_2)\prod_{i=1}^{M-1}(y_1-z_i)^{2}(y_2-z_i)^{2}\prod_{1\leq i<j\leq M-1}(z_i-z_j)^{4}
\label{alter1}
\ee

In order to show the first identity is correct, we start from the Vandermonde determinant as follows
\be
\mbox{det}
\begin{pmatrix}
1 & z_{1} & z_{1}^{2} & z_{1}^{3} & \cdots & z_{1}^{2M-1} \\
1 & \omega_{1} & \omega_{1}^{2} & \omega_{1}^{3} & \cdots & \omega_{1}^{2M-1} \\
1 & z_{2} & z_{2}^{2} & z_{2}^{3} & \cdots & z_{2}^{2M-1} \\
1 & \omega_{2} & \omega_{2}^{2} & \omega_{2}^{3} & \cdots & \omega_{2}^{2M-1} \\
\vdots & \vdots & \vdots & \vdots & \ddots & \vdots \\
1 & z_{M} & z_{M}^{2} & z_{M}^{3} & \cdots & z_{M}^{2M-1} \\
1 & \omega_{M} & \omega_{M}^{2} & \omega_{M}^{3} & \cdots & \omega_{M}^{2M-1}
\end{pmatrix}
=(-1)^{M(2M-1)}\prod_{i<j}(z_{i}-z_{j})\prod_{i<j}(\omega_{i}-\omega_{j})\prod_{i \leq j}(z_{i}-\omega_{j})\prod_{i<j}(\omega_{i}-z_{j})
\ee
Taking the derivatives of the above equation with respect to $\omega_j$ for $j=1,\cdots,M$ then set $\omega_j=z_j$, we find that the left hand side becomes
\be
\mbox{LHS}=\mbox{det}
\begin{pmatrix}
1 & z_{1} & z_{1}^{2} & z_{1}^{3} & \cdots & z_{1}^{2M-1} \\
0 & 1 & 2z_{1} & 3z_{1}^{2} & \cdots & (2M-1)z_{1}^{2M-2} \\
1 & z_{2} & z_{2}^{2} & z_{2}^{3} & \cdots & z_{2}^{2M-1} \\
0 & 1 & 2z_{2} & 3z_{2}^{2} & \cdots & (2M-1)z_{2}^{2M-2} \\
\vdots & \vdots & \vdots & \vdots & \ddots & \vdots \\
1 & z_{M} & z_{M}^{2} & z_{M}^{3} & \cdots & z_{M}^{2M-1} \\
0 & 1 & 2z_{M} & 3z_{M}^{2} & \cdots & (2M-1)z_{M}^{2M-2}
\end{pmatrix}
\ee
In the same time, the right hand side becomes
\be
\nonumber
\mbox{RHS} &=& (-1)^{M(2M-1)}\frac{\partial^{M}}{\partial\omega_{1}\partial\omega_{2}\cdots\partial\omega_{M}}\Big [\prod_{i<j}(z_{i}-z_{j})\prod_{i<j}(\omega_{i}-\omega_{j})\prod_{i \leq j}(z_{i}-\omega_{j})\prod_{i<j}(\omega_{i}-z_{j})\Big ]\Big |_{\omega_{j}\rightarrow z_{j}}
\ee
The derivatives will be applied to the 3 factors that involving $\omega_j$. Because the factor $\prod_{i<j}(\omega_{i}-\omega_{j})$ is anti-symmetric under $i\leftrightarrow j$ but the derivatives is symmetric under $i\leftrightarrow j$, we have that $\frac{\p^{M}\prod_{i<j}(\omega_{i}-\omega_{j})}{\p\omega_{1}\p\omega_{2}\cdots\p\omega_{M}}=0$. Because the factor $\prod_{i<j}(\omega_{i}-z_{j})$ do not contain $\omega_M$, we find that $\frac{\partial^{M}\prod_{i<j}(\omega_{i}-z_{j}) }{\p\omega_{1}\p\omega_{2}\cdots\p\omega_{M}}=0$. Similarly£¬  the factor $\prod_{i<j}(z_{i}-\omega_{j})$ do not contain $\omega_1$, we find that $\frac{\partial^{M}\prod_{i<j}(z_{i}-\omega_{j}) }{\p\omega_{1}\p\omega_{2}\cdots\p\omega_{M}}=0$. The only nonzero contributions come from the factor $\prod_{j}(z_{j}-\omega_{j})$, we find that
\be
\nonumber
\mbox{RHS}&=&(-1)^{M(2M-1)}\prod_{i<j}(z_{i}-z_{j})\prod_{i<j}(\omega_{i}-\omega_{j})\prod_{i<j}(\omega_{i}-z_{j})\prod_{i < j}(z_{i}-\omega_{j})\frac{\partial^{M}\prod_{j}(z_{j}-\omega_{j})}{\partial\omega_{1}\partial\omega_{2}\cdots\partial\omega_{M}}\Big |_{\omega_{j}\rightarrow z_{j}}\\
&=&(-1)^{2M^{2}}\prod_{i<j}(z_{i}-z_{j})(\omega_{i}-\omega_{j})(\omega_{i}-z_{j})(z_{i}-\omega_{j})\Big |_{\omega_{j}\rightarrow z_{j}}
=\prod_{i<j}(z_{i}-z_{j})^{4}
\ee
Therefore Eq.(\ref{alter}) is established.

To derive the second identity, we again use the following Vandermonde determinant
\be
&&\mbox{det}
\begin{pmatrix}
1 & y_1 & y_1^{2} & y_1^{3} & \cdots & y_1^{2M-1} \\
1 & y_2 & y_2^{2} & y_2^{3} & \cdots & y_2^{2M-1} \\
1 & z_{1} & z_{1}^{2} & z_{1}^{3} & \cdots & z_{1}^{2M-1} \\
1 & \omega_{1} & \omega_{1}^{2} & \omega_{1}^{3} & \cdots & \omega_{1}^{2M-1} \\
\vdots & \vdots & \vdots & \vdots & \ddots & \vdots \\
1 & z_{M-1} & z_{M-1}^{2} & z_{M-1}^{3} & \cdots & z_{M-1}^{2M-1} \\
1 & \omega_{M-1} & \omega_{M-1}^{2} & \omega_{M-1}^{3} & \cdots & \omega_{M-1}^{2M-1}
\end{pmatrix}_{2M\times 2M} \\
&&=(-1)^{2M(2M-1)/2}(y_1-y_2)\prod_{i=1}^{M-1}(y_1-\z{i})(y_1-\omega_{i})(y_2-\z{i})(y_2-\omega_{i})\\
&&\times\prod_{1\leq i<j \leq M-1}(z_{i}-z_{j})(\omega_{i}-\omega_{j})
\prod_{i\leq j}(z_{i}-\omega_{j})\prod_{i<j}(\omega_{i}-z_{j})
\ee
Taking the derivatives of the above equation with respect to $\omega_j$ for $j=1,\cdots,M-1$ then set $\omega_j=z_j$, following similar steps as in deriving the first identity, one can see that Eq.(\ref{alter1}) is correct.

\end{document}